\documentstyle[twocolumn,aps,epsf]{revtex}

\def\myfigure#1#2{{\leftskip=0.000753\textwidth
\rightskip\leftskip\small \begin{figure}\baselineskip=14pt plus 2pt
minus 1pt \centerline{#1}\nobreak\smallskip\nobreak #2\end{figure}}}
\global\firstfigfalse

\draft	

\begin{document}


\title{Arrested States of Solids}

\author{Madan Rao\cite{MAD}}

\address{Raman Research Institute, C.V. Raman Avenue,
Sadashivanagar, Bangalore 560080, 
India}

\author{and}

\author{Surajit Sengupta\cite{GAMBO}} 

\address{Material Science Division, Indira Gandhi Center for
Atomic Research, Kalpakkam 603102, India}

\maketitle

\begin{abstract}

Solids produced as a result of a fast quench across a freezing or a
structural transition get stuck in long-lived metastable configurations
of distinct morphology, sensitively dependent on the processing history.
{\it Martensites} are particularly well studied examples of nonequilibrium
solid-solid transformations. Since there are some excellent reviews on the
subject, we shall, in this brief article, mainly present our viewpoint.

\end{abstract}

\vskip 0.1 in

\section{Nonequilibrium Structures in Solids}

What determines the final microstructure of a solid under changes of
temperature or pressure ? This is an extremely complex issue, since a
rigid solid finds it difficult to flow along its free energy landscape to
settle into a unique equilibrium configuration. Solids often get stuck in
long-lived metastable or jammed states because the energy barriers that
need to be surmounted in order to get unstuck are much larger than $k_BT$.

Such nonequilibrium solid structures may be obtained either by quenching
from the liquid phase across a freezing transition (see Ref.\cite{CAROLI}
for a comprehensive review), or by cooling from the solid phase across a
structural transition. Unlike the former, nonequilibrium structures
resulting from structural transformations do not seem to have attracted
much attention amongst physicists, apart from Refs.\cite{KRUM,GOOD},
possibly because the microstructures and mechanical properties obtained
appear nongeneric and sensitively dependent on details of processing
history.

Metallurgical studies have however classified some of the more generic
nonequilibrium microstructures obtained in solid (parent/ austenite) -
solid (product/ ferrite) transformations depending on the kind of shape
change and the mobility of atoms. To cite a few :

\begin{itemize}

\item {\it Martensites} are the result of solid state transformations
involving shear and no atomic transport. Martensites occur in a wide
variety of alloys, polymeric solids and ceramics, and exhibit very
distinct plate-like structures built from twinned variants of the
product.

\item {\it Bainites} are similar to martensites, but in addition possess a
small concentration of impurities (e.g. carbon in iron) which diffuse and
preferentially dissolve in the parent phase.

\item {\it Widmanst\"atten ferrites} result from structural
transformations involving shape changes and are accompanied by short range
atomic diffusion. 

\item {\it Pearlites} are a eutectic mixture of bcc Fe and the
carbide consisting of alternating stripes.

\item {\it Amorphous} alloys, a result of a fast quench, typically possess 
some short range ordering of atoms. 

\item {\it Polycrystalline} materials of the product phase are
a result of a slower quench across a structural transition and display
macroscopic regions of ordered configurations of atoms separated by grain
boundaries. 

\end{itemize}

That the morphology of a solid depends on the detailed dynamics across a
solid-solid transformation, has been recognised by metallurgists who
routinely use time-temperature-transformation (TTT) diagrams to determine
heat treatment schedules. The TTT diagram is a family of curves
parametrized by a fraction $\delta$ of transformed product. Each curve is
a plot of the time required to obtain $\delta$ versus temperature of the
quench (Fig.\ 1). The TTT curves for an alloy of fixed composition may be
viewed as a `kinetic phase diagram'.  For example, starting from a hot
alloy at $t=0$ equilibrated above the transition temperature (upper left
corner) one could, depending on the quench rate (obtained from the slope
of a line $T(t)$), avoid the nose of the curve and go directly into the
martensitic region or obtain a mixture of ferrite and carbide when cooled
slowly.

\myfigure{\epsfysize2.2in\epsfbox{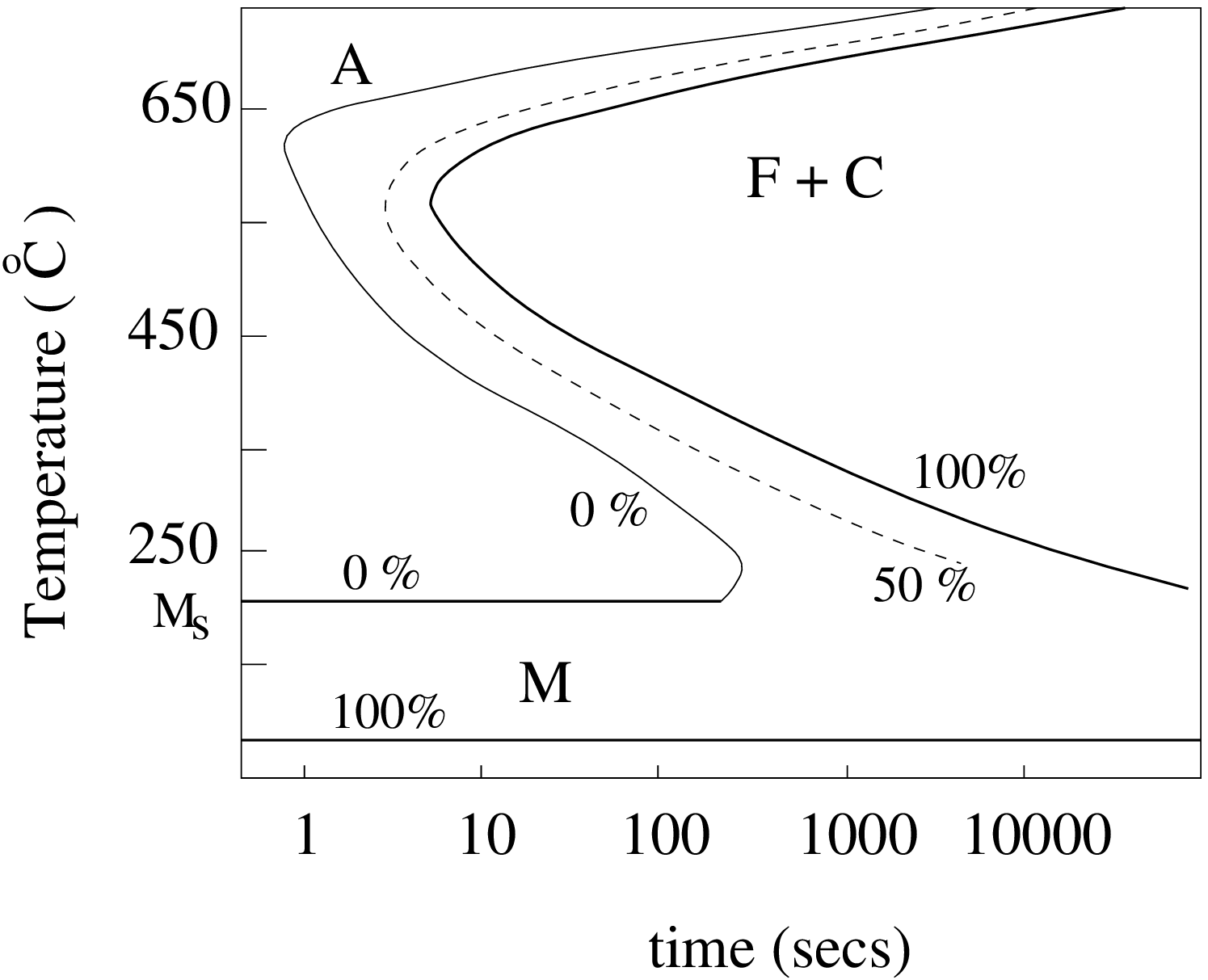}}{\vskip0inFig.\ 1~~TTT curves
\cite{MH} for steel AISI 1090 ($0.84 \%$ C + $0.60 \%$ Mn). A\,: austenite
(fcc), F\,: ferrite (bcc), C\,: carbide (Fe$_3$C).  Curves correspond to
$0$, $50$ and $100$ \% transformation. Below a temperature $M_s$, the
metastable martensite (M) is formed - the transformation curves for
martensites are horizontal.}

It appears from these studies that several qualitative features of the
kinetics and morphology of microstructures are common to a wide variety of
materials. This would suggest that there must be a set of general
principles underlying such nonequilibrium solid-solid transformations. 
Since most of the microstructures exhibit features at length scales
ranging from $100 \AA$-$100 \mu m$, it seems reasonable to describe the
phenomena at the mesoscopic scale, wherein the solid is treated as a
continuum. Such a coarse-grained description would ignore atomic details
and instead involve effective continuum theories based on symmetry
principles, conservation laws and broken symmetry. 

Let us state the general problem in its simplest context. Consider a solid
state phase diagram exhibiting two different equilibrium crystalline
phases separated by a first order boundary (Fig.\ 2). An adiabatically
slow quench from $T_{in} \to T_{fin}$ across the phase boundary in which
the cooling rate is so small that at any instant the solid
is in equilibrium corresponding to the instantaneous temperature would
clearly result in an equilibrium final product at $T_{fin}$. On the other
hand, an instantaneous quench would result in a metastable product bearing
some specific relation to the parent phase. The task is to develop a
nonequilibrium theory of solid state tranformations which would relate
the nature of the final arrested state and the dynamics leading to it to
the type of structural change, the quench rate and the mobility of atoms.

In this article we concentrate on the dynamical and structural features of
a class of solid-solid transformations called {\it Martensites}.  Because
of its commercial importance, martensitic transformations are a well
studied field in metallurgy and material science. Several classic review
articles and books discuss various aspects of martensites in great detail
\cite{NISHROIT}. The growing literature on the subject is a clear
indication that the dynamics of solid state transformations is still not
well understood. We would like to take this opportunity to present, for
discussion and criticism, our point of view on this very complex area of
nonequilibrium physics \cite{DROP,KS}.

We next review the phenomenology of martensites and highlight generic
features that need to be explained by a nonequilibrium theory of solid
state transformations.

\section{Phenomenology of Martensites}

One of the most studied alloys undergoing martensitic transformations is
iron-carbon \cite{NISHROIT}. As the temperature is reduced, Fe with less
than $0.02$\% C undergoes an equilibrium structural transition (Fig.\ 2)
from fcc (austenite) to bcc (ferrite) at $T_c = 910^{\circ}$C . An
adiabatic cooling across $T_c$ nucleates a grain of the ferrite which
grows isotropically, leading to a polycrystalline bcc solid. A faster
quench from $T_{in} > T_c$ to $T_{fin} < M_s < T_c$ (where $M_s$ \,:
martensite start temperature) produces instead a rapidly transformed
metastable phase called the martensite, preempting the formation of the
equilibrium ferrite. It is believed that martensites form by a process of
heterogeneous nucleation.  On nucleation, martensite `plates' grow
radially with a constant front velocity $\sim 10^5 cm/s$, comparable to
the speed of sound. Since the transformation is not accompanied by the
diffusion of atoms, either in the parent or the product, it is called a
diffusionless transformation. Electron microscopy reveals that each plate
consists of an alternating array of twinned or slipped bcc regions of size
$\approx 100$\AA. Such martensites are called acicular martensites. 

\myfigure{\epsfysize3.0in\epsfbox{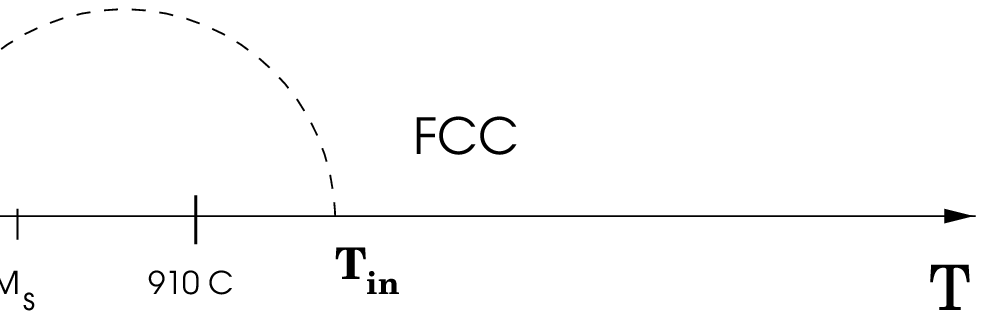}}{\vskip-1inFig.\ 
2~~Phase diagram of Fe-C (weight percent of C $< 0.02$\%). 
$M_s$ is the martensite start temperature.}

The plates grow to a size of approximately $1 \mu m$ before they collide
with other plates and stop. Most often the nucleation of plates is
athermal\,; the amount of martensite nucleated at any temperature is
independent of time. This implies that there is always some retained fcc,
partitioned by martensite plates. Optical micrographs reveal that the
jammed plates lie along specific directions known as habit planes.
Martensites, characterised by such a configuration of jammed plates, are
long lived since the elastic energy barriers for reorganisation are much
larger than $k_BT$.

A theoretical analysis of the dynamics of the martensitic transformation
in Fe-C is complicated by the fact that the deformation is 3-dimensional
(Bain strain) with 3 twin variants of the bcc phase. Alloys like In-Tl,
In-Pb, Mn-Fe and high-T$_{c}$ ceramics however, offer the simplest
examples of martensitic transformations having only two twin variants. For
instance, the high-T$_{c}$ cuprates undergo a tetragonal to orthorhombic
transformation \cite{KRUM}. The orthorhombic phase can be obtained from
the tetragonal phase by a two-dimensional deformation, essentially a
square to rhombus transition. Experiments indicate that all along the
kinetic pathway, the local configurations can be obtained from a
two-dimensional deformation of the tetragonal cell. This would imply that
the movement of atoms is strongly anisotropic and confined to the
ab-plane. Thus as far as the physics of this transformation is concerned,
the ab-planes are in perfect registry (no variation of the strain along
the c-axis). In the next two sections we shall discuss our work on the
dynamics of the square to a rhombus transformation in 2-dimensions using a
molecular dynamics simulation and a coarse-grained mode coupling theory.

\section{Molecular Dynamics Simulation of Solid-Solid Transformations}

Our aim in this section will be to study the simplest molecular dynamics
(MD) simulation of the square to rhombus transformation. We would like to
use the simulation results to construct the complete set of coarse grained
variables needed in a continuum description of the dynamics of solid state
tranformations. We carry out the MD simulation in the constant $NVT$
ensemble using a Nos\'e-Hoover thermostat ($N = 12000$) \cite{SIM}.

Our MD simulation is to be thought of as a `coarse-grained' MD simulation,
where the effective potential is a result of a complicated many-body
interaction. One part of the interaction is a purely repulsive two-body
potential $V_2(r_{ij}) = v_2/r_{ij}^{12}$ where $r_{ij}$ is the distance
between particles $i$ and $j$. The two-body interaction favours a
triangular lattice ground state. In addition, triplets of particles
interact via a short range three-body potential $V_3({\bf r}_i,{\bf
r}_j,{\bf r}_k) = v_3 w(r_{ij}, r_{jk}, r_{ik}) [ \sin^2 (4\theta_{ijk}) +
\sin^2 (4\theta_{jki})+ \sin^2 (4\theta_{kij})] $ where $w(r)$ is smooth
short-range function and $\theta_{ijk}$ is the bond angle at $j$ between
particles $(ijk)$. Since $V_3$ is minimised when $\theta_{ijk} = 0$ or
$\pi/2$, the three-body term favours a square lattice ground state. Thus
at sufficiently low temperatures, we can induce a square to triangular
lattice transformation by tuning $v_3$. The phase diagram in the $T-v_3$
plane is exhibited in Fig.\ 3.
                        
We define elastic variables, coarse-grained over a spatial block of
size $\xi$ and a time interval $\tau$, from the instantaneous positions ${\bf
u}$ of the particles. These include the deformation tensor $\partial
u_i/\partial x^k$, the full non linear strain $\epsilon_{ij}$, and the
vacancy field $\phi = \rho - {\overline \rho}$ ($\rho =$ coarse grained
local density, ${\overline \rho} =$ average density). We have kept track
of time dependence of these coarse-grained fields during the MD
simulation.

\myfigure{\epsfysize2.2in\epsfbox{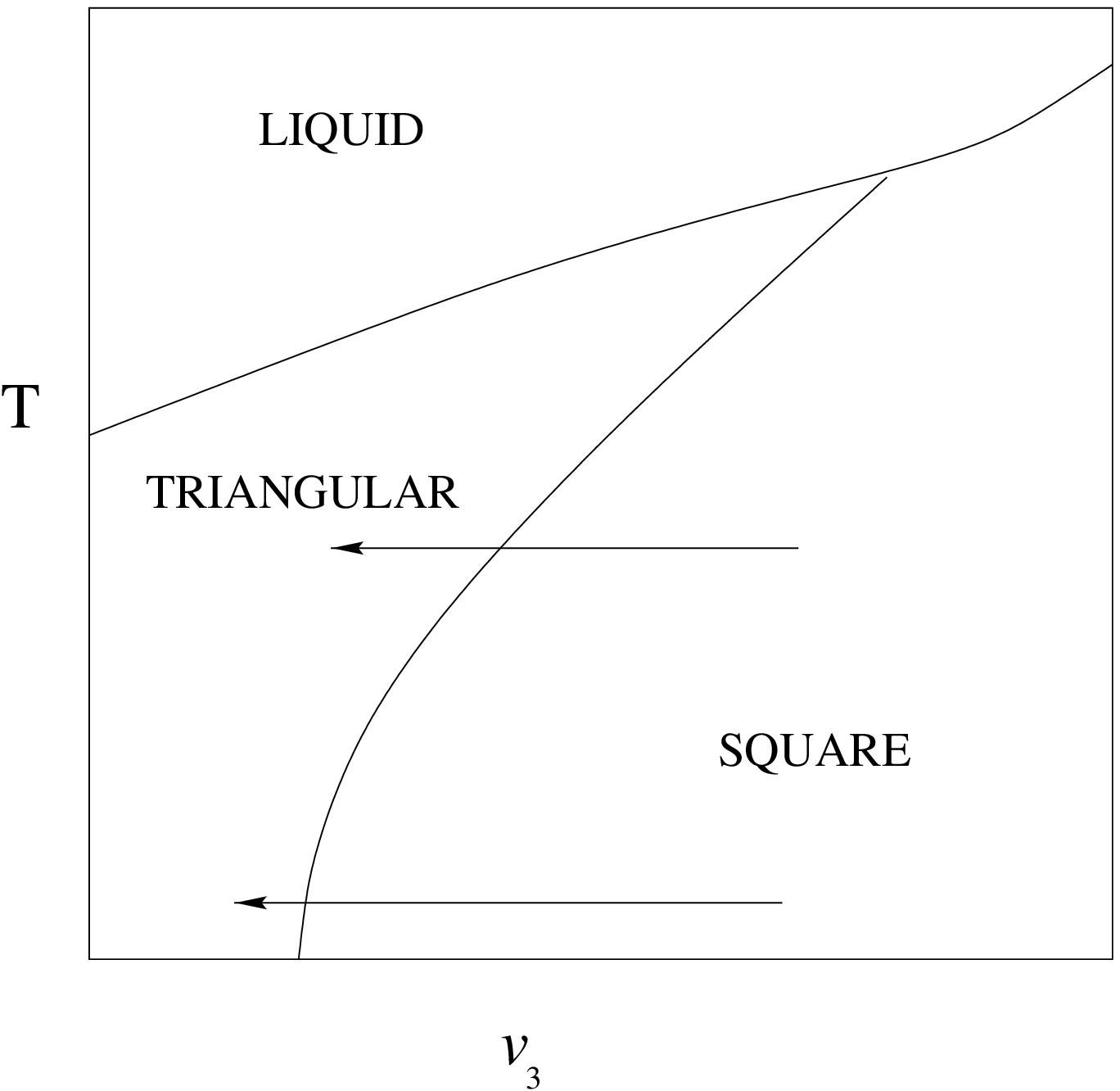}}{\vskip0inFig.\ 3~~ $T-v_3$
phase diagram from the MD simulations showing the freezing and structural
transitions. The upper and lower arrows correspond to the high and low
temperature quenches respectively.}
                              
Consider two `quench' scenarios --- a high and low temperature quench
(upper and lower arrows in Fig.\ 3 respectively) across the phase boundary.
In both cases the solid is initially at equilibrium in the square phase.

The high temperature quench across the phase boundary, induces a
homogeneous nucleation (i.e., strain inhomogeneities created by thermal
fluctuations are sufficient to induce critical nucleation) and growth of a
triangular region. The product nucleus grows isotropically with the size
$R \sim t^{1/2}$. A plot of the vacancies/interstitial field shows that,
at these temperatures they diffuse fast to their equilibrium value
(vacancy diffusion obeys an Arrenhius form $D_{v} = D_0 \exp (-A/k_BT)$,
where $A$ is an activation energy and so is larger at higher
temperatures). The final morphology is a polycrystalline triangular solid.

The low temperature quench on the other hand, needs defects (either
vacancies or dislocations) to seed nucleation in an appreciable time.
This heterogeneous nucleation initiates an embryo of triangular phase,
which grows anisotropically along specific directions (Fig.\ 4). Two
aspects are immediately apparent, the growing nucleus is twinned and the
front velocities are high. Indeed the velocity of the front is a constant
and roughly half the velocity of longitudinal sound. A plot of the
vacancy/interstitial field shows a high concentration at the
parent-product interface. The vacancy field now diffuses very slowly and
so appear to get stuck to the interface over the time scale of the
simulation. If we force the vacancies and interstitials to annihilate each
other, then the anisotropic twinned nucleus changes in the course of
time to an isotropic untwinned one !


Therefore the lessons from the MD simulation are\, : (1) There are at
least two scenarios of nucleation of a product in a parent depending on
the temperature of quench. The product grows via homogeneous nucleation at
high $T$, and via heterogeneous nucleation at low $T$ (2) The complete set
of slow variables necessary to describe the nucleation of solid-solid
transformations should include the strain tensor and defects (vacancies
and dislocations) which are generated at the parent-product interface at
the onset of nucleation (3)  The relaxation times of these defects dictate
the final morphology. At high temperatures the defects relax fast and the
grains grow isotropically with a diffusive front. The final morphology is
a polycrystalline triangular solid. At low temperatures the interfacial
defects (vacancies) created by the nucleating grain relax slowly and get
stuck at the parent-product interface. The grains grow anisotropically
along specific directions. The critical nucleus is twinned and the front
grows ballistically (with a velocity comparable to the sound speed). The
final morphology is a twinned martensite.

\section{Mode Coupling Theory of Solid-Solid Transformations}

Armed with the lessons from the MD simulation, let us now construct a
continuum elastic theory of solid-state nucleation. The analysis follows
in part the theories of Krumhansl et. al. \cite{KRUM}, but has important
points of departure.  The procedure is to define a coarse grained free
energy functional in terms of all the relevant `slow' variables. From the
simulation results, we found that every configuration is described in
terms of the local (nonsingular) strain field $\epsilon_{ij}$, the
vacancy field $\phi$, and singular defect fields like the dislocation
density $b_{ij}$. These variables are essentially related to the phase and
amplitudes of the density wave describing the solid $\{\rho_{\bf G}\}$.

It is clear from the simulation that the strain tensor, defined with
respect to the ideal parent, gets to be of $O(1)$ in the interfacial
region between the parent and the product. Thus we need to use the full
nonlinear strain tensor $\epsilon_{ij} = (\partial_i u_j + \partial_j u_i
+ \partial_i u_k \partial_j u_k)/2$. Further, since the strain is
inhomogeneous during the nucleation process, the free energy functional
should have derivatives of the strain tensor $\partial_k \epsilon_{ij}$
(this has unfortunately been termed `nonlocal strain' by some authors).

In general, the form of the free energy functional can be very
complicated, but in the context of the square-to-rhombus transformation,
the free energy density may be approximated by a simple form,

\begin{equation}
f = c\,(\nabla \epsilon)^2 + \epsilon^2 - a \epsilon^4 +
\epsilon^6 + \chi_{v} \,\phi^2 + \chi_{d} \,b^2 + k_{d}\, b\, \epsilon 
\label{eq:landau}
\end{equation}
where $\epsilon$ is the nonzero component of the strain corresponding to
the transformation between a square and a rhombus, $\phi$ is the vacancy
field and $b$ is the dislocation density (we have dropped the tensor
indices for convenience). The tuning parameter $a$ induces a transition
from a square (described by $\epsilon = 0$) to a rhombus ($\epsilon = \pm
e_0$).

Starting with $\epsilon=0$ corresponding to the equilibrium square parent
phase at a temperature $T > T_c$, we quench across the structural
transition. The initial configuration of $\epsilon$ is now metastable at
this lower temperature, and would decay towards the true equilibrium
configuration by nucleating a small `droplet' of the product. As we saw in
the last section, as soon as a droplet of the product appears embedded in
the parent matrix, atomic mismatch at the parent-product interface gives
rise to interfacial defects like vacancies and dislocations.  

Let us confine ourselves to solids for which the energy cost of producing
dislocations is prohibitively large. This would imply that the interfacial
defects consist of only vacancies and interstitials. The dynamics of
nucleation now written in terms of $\epsilon$, $\bf g$ (the conserved
momentum density) and vacancy $\phi$ are complicated \cite{DROP}. For the
present purpose, all we need to realise is that $\phi$ couples to the
strain and is diffusive with a diffusion coefficient $D_{v}$ depending
on temperature.

As in the MD simulation, we find that the morphology and growth of the
droplet of the product depends critically on the diffusion of these
vacancies. If the temperature of quench is high, $\phi$ diffuses to zero
before the critical nucleus size is attained and the nucleus eventually
grows into an equilibrium (or polycrystalline) triangular solid. In this
case, the nucleus grows isotropically with $R \sim t^{1/2}$. However
a quench to lower temperatures results in a low vacancy diffusion
coefficient. In the limit $D_{v} \to 0$, the $\phi$-field remains
frozen at the moving parent-product interface. In this case a constrained
variational calculation of the morphology of the nucleus, shows that it is
energetically favourable to form a twinned martensite
rather than a uniform triangular structure. The growth of the
twinned nucleus is not isotropic, but along habit planes. Lastly the
growth along the longer direction is ballistic with a velocity
proportional to $\sqrt {\chi_{v}}$ (of the order of the sound
velocity). All these results are consistent with the results of the
previous section and with martensite phenomenology. Let us try and
understand in more physical terms, why the growing nucleus might want to
form twins.

As soon as a droplet of the triangular phase of dimension $L$ is
nucleated, it creates vacancies at the parent-product interface. The free
energy of such an inclusion is $F = F_{bulk} + F_{pp} + F_{\phi}$. The
first term is simply the bulk free energy gain equal to $\Delta F L^2$
where $\Delta F$ is the free energy difference between the square and
triangular phases. The next two terms are interfacial terms. $F_{pp}$ is
the elastic contribution to the parent-product interface coming from the
gradient terms in the free energy density Eq.\ \ref{eq:landau}, and is
equal to $4 \sigma_{pp} L$, where $\sigma_{pp}$ is the surface tension at
the parent-product interface. $F_{\phi}$ is the contribution from the
interfacial vacancy field glued to the parent-product interface and is
proportional to $\phi^2 \sim L^2$ (since the atomic mismatch should scale
with the amount of parent-product interface). This last contribution
dominates at large $L$ setting a prohibitive price to the growth of the
triangular nucleus. The solid gets around this by nucleating a twin with a
strain opposite to the one initially nucleated, thereby reducing $\phi$.
Indeed for an equal size twin, $\phi \to 0$ on the average, and leads to a
much lower interfacial energy $F_{\phi} \sim L$. However the solid now
pays the price of having created an additional twin interface whose energy
cost is $F_{tw} = \sigma_{tw} L$.

Considering now an (in general) anisotropic inclusion of length $L$, width
$W$ consisting of  $N$ twins, the free energy calculation goes as
\begin{equation} 
F = {\Delta F} L W + \sigma_{pp} (L+W) + \sigma_{tw} N W
+ \beta \left(\frac{L}{N}\right)^2 N 
\end{equation} 
where the last term is the vacancy contribution. Minimization with respect
to $N$ gives $L/N \sim W^{1/2}$, a relation that is known for
2-dimensional martensites like In-Tl.

Our next task is to solve the coupled dynamical equations with appropriate
initial conditions numerically, to obtain the full morphology phase
diagram as a function of the type of structural change, the parameters
entering the free energy functional and kinetic parameters like $D_{v}$.

It should be mentioned that our theory takes off from the theories of
Krumhansl et. al. \cite{KRUM}, in that we write the elastic energy in
terms of the nonlinear strain tensor and its derivatives. In addition we
have shown that the process of creating a solid nucleus in a parent
generates interfacial defects which evolve in time. The importance of
defects has been stressed by a few metallurgists \cite{MET}. We note also
that the parent-product interface is studded with an array of vacancies
with a separation equal to the twin size. This implies that the strain
decays exponentially from interface over a distance of order $L/N$. This
has been called `fringing field' in Ref.\ \cite{KRUM}. Krumhansl et. al.
obtain this by imposing boundary conditions on the parent-product
interface, whereas here it appears dynamically.

\section{Patterning in Solid-Solid Tranformations\,: Growth and Arrest}

So far we have discussed the nucleation and growth of single grains. This
description is clearly valid at very early times, for as time progresses
the grains grow to a size of approximately $1 \mu m$ and start colliding,
whereupon in most alloys they stop. Optical micrographs of acicular
martensites reveal that the jammed plates lie along habit planes that
criss-cross and partition the surrounding fcc (parent) matrix.

Can we quantify the patterning seen in martensite aggregates over a scale
of a millimeter ? A useful measure is the size distribution of the
martensite grains embedded in a given volume of the parent. The
appropriate (but difficult !) calculation at this stage would
be the analogue of a Becker-D\"oring theory for nucleation in solids.
In the absence of such a theory, we shall take a phenomenological
approach.

Clearly the size distribution $P(l,t)$ depends on the spatio-temporal
distribution $I$ of nucleation sites and the growth velocity $v$. We have
analysed the problem explicitly in a simple 2-dimensional context. Since
the nucleating martensitic grains are highly anisotropic and grow along
certain directions with a uniform velocity, a good approximation is to
treat the grains as lines or rays. These rays (lines) emanate from
nucleation sites along certain directions, and grow with a constant
velocity $v$. The rays stop on meeting other rays and eventually after a
time $T$, the 2-dimensional space is fragmented by $N$ colliding rays. The
size distribution of rays, expressed in terms of a scaling variable $y =
y(I,v)$, has two geometrical limits --- $\bf {\Gamma}$-fixed point (at
$y=0$) and the $\bf L$-fixed point (at $y=\infty$). The $\bf
{\Gamma}$-fixed point corresponds to the limit where the rays nucleate
simultaneously with a uniform spatial distribution. The stationary
distribution $P(l)$ is a Gamma distribution with an exponentially decaying
tail. The $\bf L$-fixed point, corresponds to the limit where the rays are
nucleated sequentially in time (and uniformly in space) and grows with
infinite velocity.  By a mapping onto a multifragmentation problem, Ben
Naim and Krapivsky \cite{NK} were able to derive the exact asymptotic form
for the moments of $P(l)$ at the {\bf L}-fixed point. The distribution
function $P(l)$ has a multiscaling form, characterised by its moments
$<l^q> \sim N^{-\mu(q)}$ where $\mu(q) = (q+2-\sqrt{q^2+4})/2$. At
intermediate values of the scaling variable $y$, there is a smooth
crossover from the $\bf {\Gamma}$-fixed point to the $\bf L$-fixed point
with a kinematical crossover function and crossover exponents.

The emergence of scale invariant microstructures in martensites as arising
out of a competition between the nucleation rate and growth is a novel
feature well worth experimental investigation. There have been similar
suggestion in the literature, but as far as we know there has been no
direct visualization studies of the microstructure of acicular martensites
using optical micrographs. Recent acoustic emission experiments
\cite{VIVES} on the thermoelastic reversible martensite Cu-Zn-Al, may be
argued to provide indirect support of the above claim \cite{DROP}, but the
theory of acoustic emission in martensites is not understood well enough
to make such an assertion with any confidence.

\section{Open Questions}

We hope this short review makes clear how far we are in our understanding
of the dynamics of solid-solid transformations. A deeper understanding of
the field will only come about with systematic experiments on carefully
selected systems. For instance, a crucial feature of our nonequilibrium
theory of martensitic transformations is the existence of a dynamical
interfacial defect field. In conventional Fe based alloys, the martensitic
front grows incredibly fast, making it difficult to test this using {\it
in situ} transmission electron microscopy. Colloidal solutions of
polysterene spheres (polyballs) however, are excellent systems for
studying materials properties. Polyballs exhibiting fcc $\to$ bcc
structural transitions have been seen to undergo twinned martensitic
transformations. The length and time scales associated with colloids are
large, making it comfortable to study these systems using light scattering
and optical microscopy.

In this article we have focussed on a small part of the dynamics of solid
state transformations, namely the dynamics and morphology of martensites.
Even so our presentation here is far from complete and there are crucial
unresolved questions that we need to address.

Let us list the issues as they appear following a nucleation event.  

The physics of heterogeneous nucleation in solids is very poorly
understood. For instance, it appears from our simulations that the
morphology of the growing nucleus depends on the nature of the defects
seeding the nucleation process (e.g., vacancies, dislocations and grain
boundaries). In addition several martensitic transformations are
associated with correlated nucleation events and autocatalysis.  Though
these features are not central to the issue of martensites, such a study
would lead to a better understanding of the origins of athermal,
isothermal and burst nucleation. This in conjunction with a
`Becker-D\"oring theory' for multiple martensite grains would be a first
step towards a computation of the TTT curves.

We still do not understand the details of the dynamics of twinning and how
subsequent twins are added to the growing nucleus. Moreover the structure
and dynamics of the parent-product interface and of the defects embedded
in it have not been clearly analysed.

It would be desirable to have a more complete theory which displays a
morphology phase diagram (for a single nucleus) given the type of
structural transition and the kinetic, thermal and elastic parameters.

Certain new directions immediately suggest themselves. For instance, the
role of carbon in interstitial alloys like Fe-C leading to the formation
of {\it bainites}\,; the coupling of the strain field to an external
stress and the associated {\it shape memory} effect\,; and finally the
nature of {tweed} phases and pre-martensitic phenomena (associated with
the presence of quenched impurities).

It is clear that the study of the dynamics of solid-solid transformations
and the resulting long-lived morphologies lies at the intersection of
metallurgy, material science and nonequilibrium statistical mechanics. The
diversity and richness of phenomena make this an extremely challenging
area of nonequilibrium physics.

\section{Acknowledgements}

We thank Yashodhan Hatwalne for a critical reading of the manuscript.

\newpage

\begin{center}
{\bf Figure Caption}
\end{center}

Fig.\ 4~~ MD snapshot of (a) the nucleating grain at some intermediate
time initiated by the low temperature quench across the square-triangle
transition. The dark(white) region is the triangular(square) phase
respectively. Notice that the nucleus is twinned and highly anisotropic.
(b) the vacancy (white)/interstitial (black) density profile at the same
time as (a). Notice that the vacancies and interstitials are well
separated and cluster around the parent-product interface.

\end{document}